\documentclass[twocolumn]{aa}
\usepackage{graphicx}
\usepackage{txfonts}
\usepackage{natbib}

\newcommand{\mbf}[1]{\ensuremath\mbox{\boldmath{$#1$}}}
\newcommand{\req}[1]{(\ref{#1})}

\begin{document}

  \title{The Cellular Burning Regime in Type Ia Supernova Explosions}

  \subtitle{ I.~Flame Propagation into Quiescent Fuel}


  \titlerunning{Cellular Burning in SNe Ia---Flame
                Propagation into Quiescent Fuel}

  \author{F. K. R{\"o}pke\inst{1},  W. Hillebrandt\inst{2}, \and
          J. C. Niemeyer\inst{3}}

  \authorrunning{F.~K.~R{\"o}pke et al.}

   \offprints{F. K. R{\"o}pke}

   \institute{Max-Planck-Institut f\"ur Astrophysik,
              Karl-Schwarzschild-Str. 1, D-85741 Garching, Germany\\
              \email{fritz@mpa-garching.mpg.de}
              \and
              Max-Planck-Institut f\"ur Astrophysik,
              Karl-Schwarzschild-Str. 1, D-85741 Garching, Germany\\
              \email{wfh@mpa-garching.mpg.de}
              \and
              Universit\"at W\"urzburg,
              Am Hubland, D-97074 W\"urzburg, Germany \\
              \email{niemeyer@astro.uni-wuerzburg.de}
             }

  \abstract{We present a numerical investigation of the cellular burning
    regime in Type Ia supernova explosions. This regime holds at small
    scales (i.e.~below the Gibson scale), which are unresolved in
    large-scale Type Ia supernova simulations. The fundamental effects that
    dominate the flame evolution here are the Landau-Darrieus instability
    and its nonlinear stabilization, leading to a stabilization of the
    flame in a cellular shape. The flame propagation into quiescent fuel
    is investigated addressing the dependence of the simulation
    results on the specific parameters of the numerical
    setup. Furthermore, we investigate the flame stability at a range of fuel
    densities. This is directly connected to the questions of active
    turbulent combustion (a mechanism of flame destabilization and
    subsequent self-turbulization) and a deflagration-to-detonation transition of
    the flame. In our simulations we find no substantial destabilization
    of the flame when propagating into quiescent fuels of densities down
    to $\sim 10^7 \,\mathrm{g} \,\mathrm{cm}^{-3}$, corroborating fundamental
    assumptions of large-scale SN Ia explosion models. For these models,
    however, we suggest an increased lower cutoff for the flame
    propagation velocity to take the cellular burning regime into
    account.

  }

  \maketitle

\section{Introduction}

Type Ia supernovae (SNe Ia) are usually attributed to
thermonuclear explosions of  white dwarf (WD) stars consisting of carbon and
oxygen. Throughout this paper, we will refer to the currently favored
model in which the combustion proceeds in form of a flame starting near the
center of the WD and traveling outward. More specifically, the
initial mode of flame propagation is assumed to be the so-called
deflagration mechanism, in which the combustion wave is mediated by
microphysical transport processes giving rise to a subsonic flame
speed. Underlying our assumptions is the canonical
single-degenerate Chandrasekhar-mass SN Ia 
model. For a review of SN Ia explosion models we refer to
\citet{hillebrandt2000a}. 

Despite the success of recent attempts to numerically model Type Ia
supernova explosions \citep{reinecke2002d, gamezo2003a}, some
basic questions regarding the explosion mechanism remain
unanswered. This is mainly due to the fact that the whole range of scales relevant to this
problem cannot be resolved in a single simulation in the foreseeable
future. From the radius of the exploding white dwarf star down to the flame
thickness it covers more than 11 orders of magnitude. Therefore,
large-scale simulations provide an insight into the flame dynamics on the
largest scales but have to rely on assumptions about the physics of flame
propagation on smaller scales, which is poorly understood so
far. However, the
motivation for this approach is that the key feature of the
SN Ia explosion is turbulent combustion. Turbulence is driven from large-scale
instabilities---mainly the buoyant Rayleigh-Taylor instability and the
Kelvin-Helmholtz (shear) instability---that give rise to the formation
of a turbulent eddy cascade. The turbulent eddies wrinkle the flame
and increase its surface. This enhances the net burning rate and
accelerates the flame. In this way the energy production rate can
reach values that are sufficient to power a SN Ia explosion. The underlying
assumption of this model is that the flame evolution is dominated by
the turbulent cascade originating from instabilities on large scales.
Although there exist some theoretical
ideas corroborating the assumption of flame stability on small
scales, a justification by means of a
hydrodynamical simulation has not provided definitive answers yet
\citep{niemeyer1995a, niemeyer1997b}. The
results that will be reported in the present paper are, however, a
contribution into this direction.

The work presented in the following stands in direct succession of
\citet{roepke2003a}, where the basic method was introduced and applied
to describe flame propagation into quiescent fuel for a density of the
unburnt material of $\rho_u= 5 \times 10^7 \,\mathrm{g} \,\mathrm{cm}^{-3}$. This
example demonstrated that the applied method appropriately models the
flame propagation on specific scales in the SN Ia explosion. Moreover, it
confirmed for the first time by means of a full hydrodynamical
simulation that the stabilization of the flame front in a cellular
pattern known from chemical flames holds for thermonuclear flames
under the conditions of SN Ia explosions.
In the present paper
we will extend the study of flame propagation to a wider range of fuel
densities. The aim is to test the stability of the cellular pattern
in dependence of this parameter. This will provide an
overview over possible effects of
flame propagation into quiescent fuel. We will additionally investigate the
flow field resulting from the flame configuration in greater detail
and measure the effective flame  propagation velocities for
the resulting flame structures.

The theoretical context of this study will be introduced in
Sect.~\ref{backgr_sec}. After a brief description of the applied
numerical methods in Sect.~\ref{setup_sec}, we will present the results
of numerical simulations of the flame evolution in
Sects.~\ref{quif_general_sec} and \ref{quif_fuel_sec}. Finally,
conclusions regarding the significance of the numerical investigations
for large-scale SN Ia models will be drawn.

\section{Theoretical background}
\label{backgr_sec}

Our study of flame evolution focuses on scales where flame
propagation decouples from the turbulent cascade. On larger scales,
this cascade
dominates flame propagation by wrinkling the flame and enhancing its
surface. But
in order to contribute to the flame wrinkling,
a turbulent eddy has to deform the flame significantly in a time
shorter than the flame crossing time over the spatial extent of that
eddy. Thus, equating the eddy
turnover time to this time scale, one obtains the cutoff scale
$l_\mathrm{Gibs}$
\begin{equation}\label{Gibs_eq}
v'(l_\mathrm{Gibs}) = s_l,
\end{equation}
where $s_l$ denotes the so-called laminar burning velocity of a
deflagration flame, i.e.~the propagation speed of a planar
deflagration wave which is determined by microphysical transport. 
The turbulent velocity fluctuations at length scale $l$ are
represented by $v'(l)$. 
Equation~\req{Gibs_eq} defines a length scale $l_\mathrm{Gibs}$, which
was identified with the
\emph{Gibson scale} for chemical flames \citep{peters1986a}. We will
make use of this term in the context of thermonuclear flames in SN Ia 
explosions, too.

In our model, we describe the flame as a discontinuity between burnt
and unburnt states ignoring any internal structure. This
simplification is exploited by the numerical method of flame
tracking applied in our simulations, resulting in an increased
dynamical range compared to models that fully resolve the flame
structure. 
In order to justify our \emph{thin flame approximation}, we
provide a rough estimate of the Gibson scale.
The turbulent velocity fluctuations $v'$ in eq.~\req{Gibs_eq} are
given by the scaling law of the turbulent eddy cascade. Kolmogorov
scaling, $v'(l) \propto l^{1/3}$, provides a reasonable
approximation. 
Hence
one obtains from \req{Gibs_eq}
\begin{equation}
l_\mathrm{Gibs}=L\left( \frac{s_l}{v'(L)} \right)^3.
\end{equation}
The large scale instabilities give rise to the formation of ascending
bubbles. A typical size of a Rayleigh-Taylor bubble is $L \sim 10^7
\,\mathrm{cm}$, which defines the integral scale of the turbulence.
The rise of these bubbles invokes  shear velocities $v'(L)$ of about $10^7
\,\mathrm{cm} \,\mathrm{s}^{-1}$.
Values for the laminar burning velocity have been determined by means
of one-dimensional simulations resolving the flame structure by
\citet{timmes1992a}. The authors also provide the following fitting formula to the
data which we will use in our simulations:
\begin{equation}\label{TW_eq}
s_l = 92.0\times 10^5 \left( \frac{\rho_u}{2 \times 10^9} \right)^{0.805}\left[
  \frac{X(^{12}\mathrm{C})}{0.5}\right]^{0.889} \,\mathrm{cm} \,\mathrm{s}^{-1}.
\end{equation}
However, we note that the
fit does not reproduce the simulation results as accurately as the
authors claim for the fuel density range that will be investigated in
the following. According to this formula, the laminar burning velocity
at a fuel density of $5 \times 10^7 \,\mathrm{g} \,\mathrm{cm}^{-3}$ is of the
order of $10^6 \,\mathrm{cm} \,\mathrm{s}^{-1}$. This yields a Gibson scale of
$\sim$$10^4 \,\mathrm{cm}$.
Thus, keeping in mind the indicative
character of our study, we will arrange our simulations at length
scales around this value, which is still well-separated from the flame
width.

Below the Gibson scale, flame propagagation does not
interact with the turbulent cascade, as in the context of SN Ia
explosion was first pointed out by \citet{niemeyer1995a}.
Here, it would  propagate as a laminar
flame. However, it is well-known that laminar flames are subject to a
hydrodynamical instability \citep{darrieus1938a,landau1944a}. 
The origin of this so-called \emph{Landau-Darrieus instability} (LD
instability henceforth) is
the refraction of the streamlines of the flow on the density contrast across
the flame.
Mass flux conservation leads to a broadening of the flow
tubes in the vicinity of bulges of the perturbation. 
Here the local fluid
velocity decreases and becomes lower than the fluid velocities at $\pm
\infty$. These velocities correspond to the laminar burning speeds
with respect to fuel and ashes in the frame of reference comoving with
the flame.
Therefore 
the burning velocity $s_l$ of the flame is higher than the
corresponding local fluid velocity and this leads to an increment of the
bulge. The opposite holds for recesses of the perturbed front. In this
way the perturbation keeps growing. 

\begin{figure}[ht]
\centerline{
\includegraphics[width=0.8
\hsize]{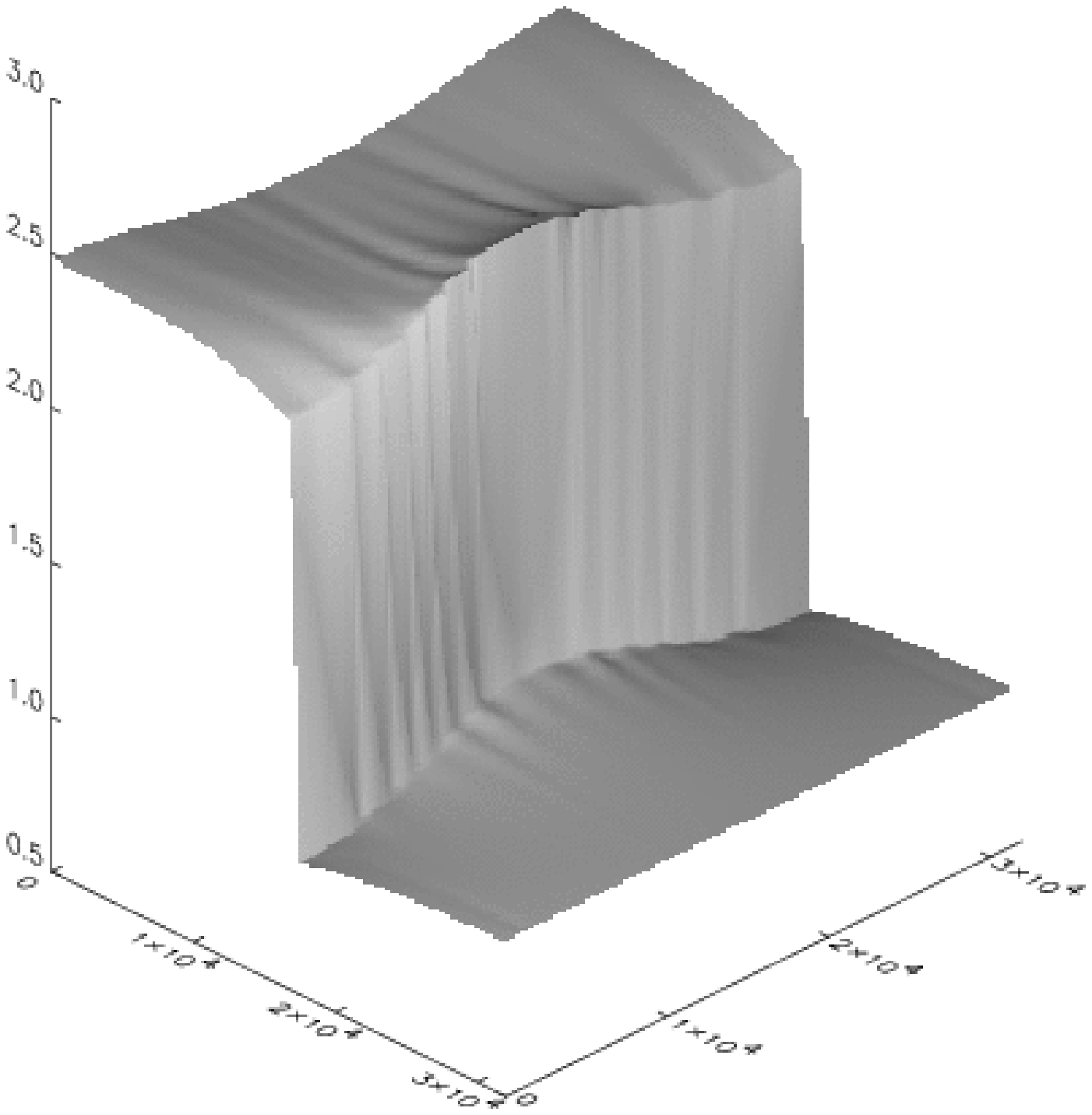}{\small(a)}}
\centerline{
\includegraphics[width=0.8 \hsize]{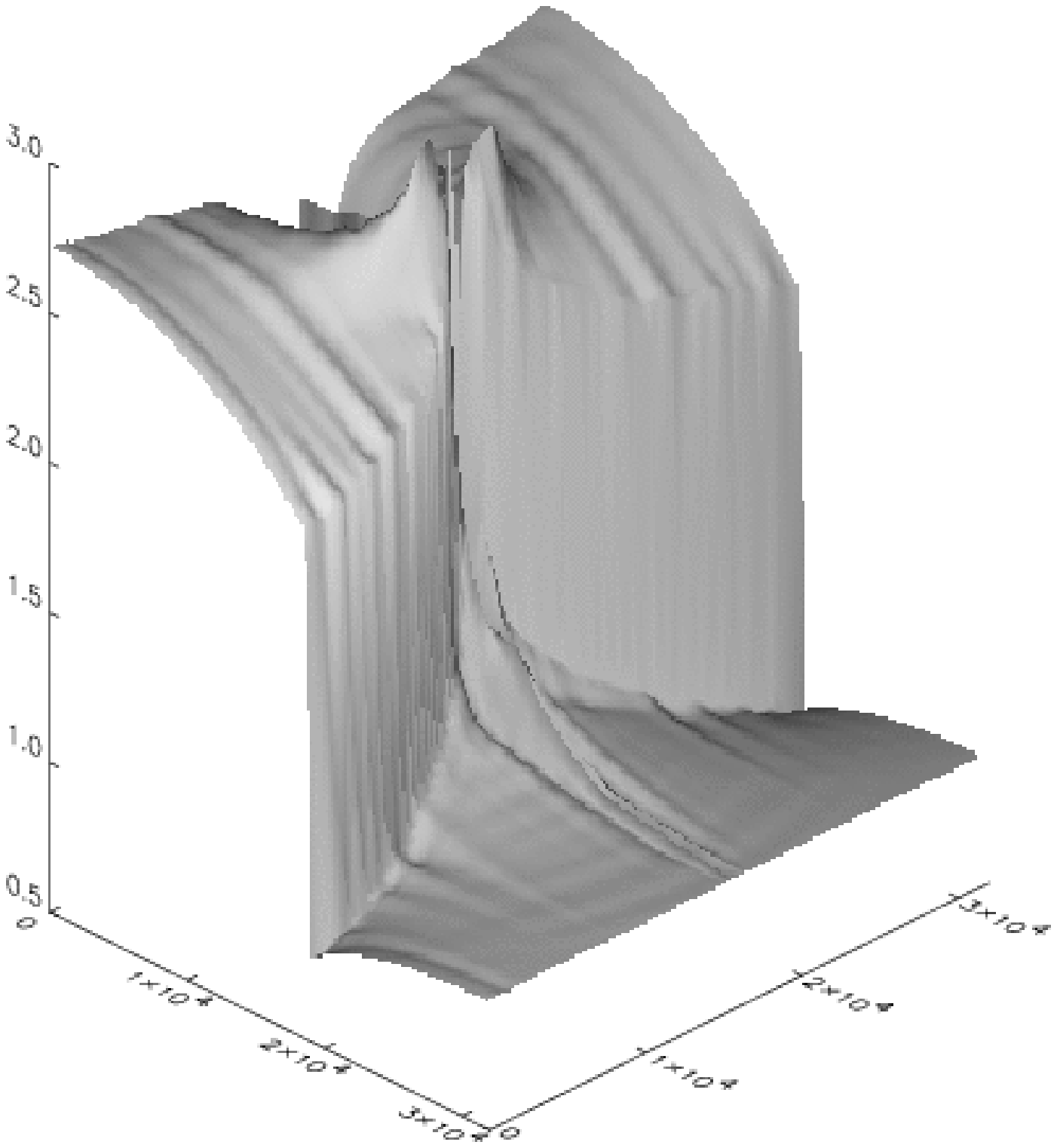}{\small(b)}}
\caption{Fluid velocity in the simulation with $\rho_u = 5 \times 10^7
  \,\mathrm{g} \,\mathrm{cm}^{-3}$ and 
  $100 \times 100$ cells resolution. The $x$-axis and the $y$-axis show position in
  cm; the $z$-axis represents $|\mbf{v}|/s_l$ \textbf{(a)} snapshot at $t=0.01
  \,\mathrm{s}$, \textbf{(b)} snapshot at $t=0.09 \,\mathrm{s}$
  \label{quif_velsurf_fig}}
\end{figure}

This can be observed in our simulations.
Figure
\ref{quif_velsurf_fig}a
depicts the absolute value of the fluid velocity normalized to the
laminar burning velocity of the flame in
the linear regime of flame evolution in one of our simulations.
It is well in agreement with the theoretical expectations and
confirmes that our numerical
method is capable to reproduce the LD instability. This is not trivial
as can be seen from the failure of the passive implementation to
develop the correct flow field (cf. \citealt{roepke2003a}).

By means of a linear stability
analysis, \cite{landau1944a} derived a dispersion relation between the
growth rate of the amplitude $\omega$ and the wavenumber $k$ of the
perturbation:
\begin{equation}
\label{omega_ld}
\omega_\mathrm{LD} = k s_l \frac{\mu}{1 + \mu}\left(\sqrt{1 + \mu -
    \frac{1}{\mu}}-1\right),
\end{equation} 
with $\mu = \rho_u/\rho_b$.

This instability,
however, is not necessarily a contradiction to the assumption of flame
stability on small scales. In fact, one observes stable flame propagation in
chemical combustion experiments. The reason for this is that the flame stabilizes in a
cellular pattern as soon as the growth of the perturbation enters the
nonlinear regime. This effect has been explained by \citet{zeldovich1966a}
in a geometrical approach. The
propagation of an initially 
sinusoidally perturbed flame is followed by means of Huygens'
principle (see also Fig.~1 in \citealt{roepke2003a}). At the intersections of two bulges of the front a cusp
forms after a while. Here Huygens' principle is no longer applicable
and the flame evolution enters the nonlinear regime. The propagation
velocity at the cusp exceeds $s_l$:
\begin{equation}
v_\mathrm{cusp} = \frac{s_l}{\cos \theta},
\end{equation}
where $\theta$ denotes the inclination angle of the front adjacent to
the cusp.
This effect balances the perturbation growth due to the LD instability
and leads to a stabilization of the flame in a cellular
configuration. Figure \ref{quif_velsurf_fig}b shows the absolute value of the
velocity after a cusp has formed.
 
Assuming a parabolic shape of the cells, the temporal evolution
of the  amplitude of the perturbation $\hat{x}$ can now be written as \citep{zeldovich1966a} 
\begin{equation}\label{zel1}
\frac{\mathrm{d} \hat{x}}{\mathrm{d} t} = \omega_\mathrm{LD} \hat{x} -  \frac{2}{\pi^2} k^2 s_l
\hat{x}^2. 
\end{equation}
The quadratic damping term accounting for the cellular stabilization
introduces an extension to the linear stability 
analysis resulting in the dispersion relation \req{omega_ld}. If one
further assumes that the flame front advances
with the constant laminar
burning speed $s_l$ independently of the flame shape, the effective 
speed $v_\mathrm{eff}$ of the mean
position of a flame wrinkled due to instabilities is given in terms
of its surface $A'$ compared to the surface of a planar flame $A(0)$:
\begin{equation}\label{ueff_a_eq}
v_\mathrm{eff} = s_l \frac{A'}{A_0}.
\end{equation}

The flame thus accelerates with increasing surface area and the task
to determine the effective flame propagation velocity reduces to the
determination of the increase in flame surface area.
From the assumption of a parabolic cell shape \citet{zeldovich1980a}
derived the acceleration of the flame corresponding to the increase in
its surface:
\begin{equation}\label{vel_inc_eq}
\frac{v_\mathrm{cell}}{s_l} = 1+ \epsilon, \qquad \epsilon =
\frac{\pi^2}{24} \left(1-\frac{1}{\mu}\right)^2.
\end{equation}

The cellular stabilization is the reason that large-scale SN Ia
simulations assume flame stability on small scales. This is, however,
not yet well confirmed. The cellular pattern may break up under
certain conditions, e.g. density jumps over the flame front or
interaction with turbulence.  \citet{blinnikov1996a} suspected that a
fractalization of
the cellular flame may drastically accelerate the flame at low
fuel densities and \citet{niemeyer1995a} reported on
indications that the flame may destabilize at
fuel densities of $\rho_u \approx 5 \times 10^7 \,\mathrm{g} \,\mathrm{cm}^{-3}$.
 However, a failure of the cellular flame stabilization
at $\rho_u = 5 \times 10^7 \,\mathrm{g} \,\mathrm{cm}^{-3}$
could not be confirmed by \citet{roepke2003a}. Nevertheless, the
possible loss of flame stability at certain densities of unburnt
material could have drastic impact on large-scale SN Ia models and has
to be soundly tested. Besides the break-down of one fundamental
assumption of these models, it could also have effects that seem
appealing. Spectra produced by empirical one-dimensional SN Ia models
fit the observations remarkably well if a transition of the flame
propagation mode from subsonic deflagration to supersonic detonation
is applied at low densities \citep{hoeflich1996a,iwamoto1999a}. The
drawback of these \emph{delayed detonation models} is that the
mechanism providing the transition is unclear
\citep{niemeyer1999a}. As one possibility, an active production of turbulence and
self-turbulization of the flame after a break-down of its cellular
stabilization was proposed under the name \emph{active turbulent
  combustion} by \citet{niemeyer1997b} based on an idea by
\citet{kerstein1996a}. Following the reasoning provided by
\citet{niemeyer1997b}, this effect is directly connected to the
burning regime we are aiming at in our simulations.

The above considerations point out the need for a thorough
investigation of the cellular burning regime in SN Ia explosions
at scales around or below the Gibson length. Numerical simulations
addressing this question will be presented in
Sects.~\ref{quif_general_sec} and \ref{quif_fuel_sec}.

\section{Numerical methods and simulation setup}
\label{setup_sec}

\begin{figure*}[p!]
\centerline{
\includegraphics[height=0.93 \textheight]{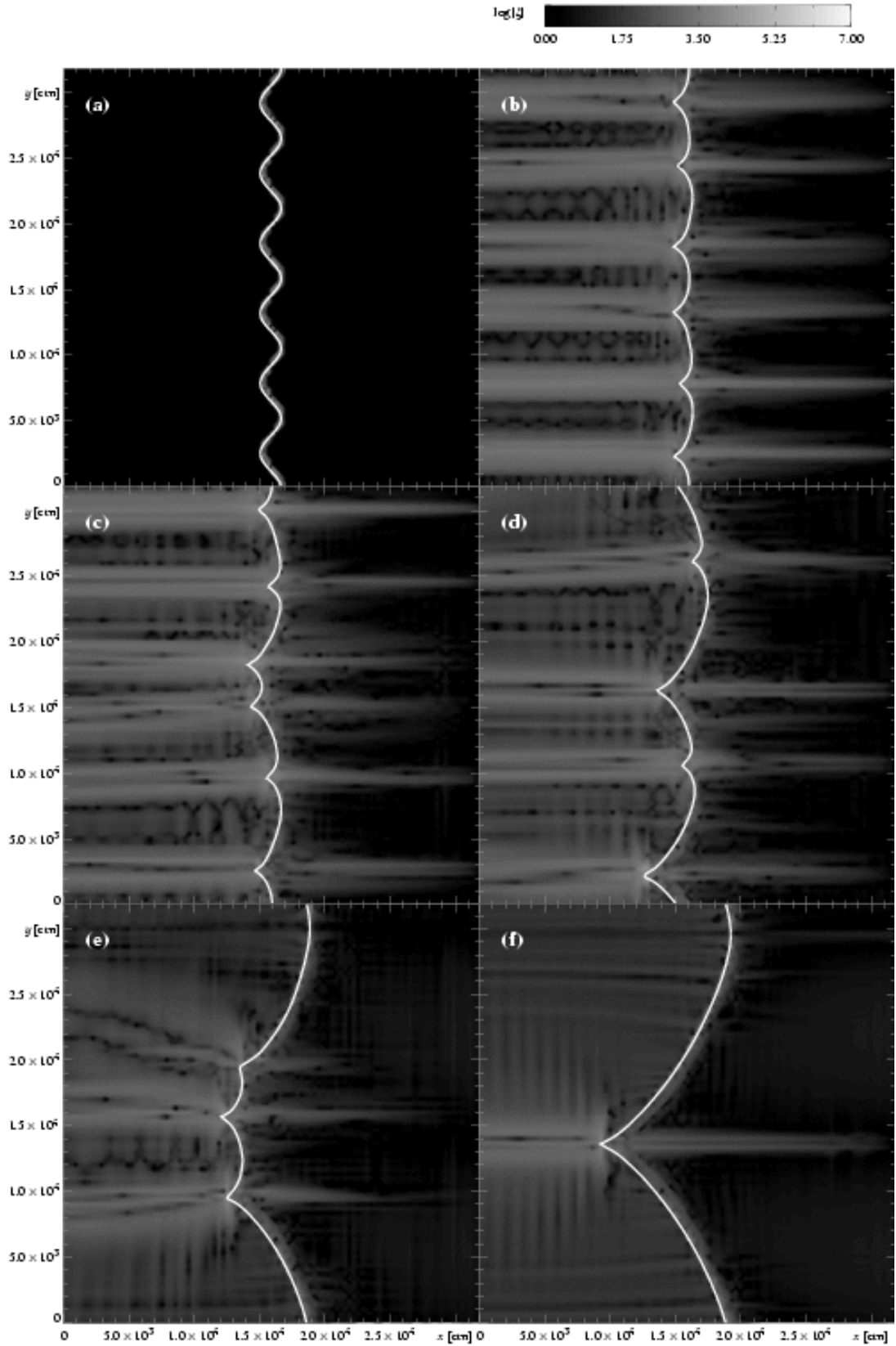}}
\caption{Flame propagation into quiescent fuel at $\rho_u = 5 \times
  10^7 \,\mathrm{g} \,\mathrm{cm}^{-3}$; resolution: $200 \times 200$ cells,
  \emph{periodic} boundary conditions; Snapshots taken at
\textbf{(a)} 0, 
\textbf{(b)} 5,
\textbf{(c)} 7.5, 
\textbf{(d)} 10,
\textbf{(e)} 15, and 
\textbf{(f)} 30 growth times
  $\tau_\mathrm{LD} = \omega_\mathrm{LD}^{-1}$ of a
  perturbation with $\lambda=200 \Delta x$. The vorticity $\zeta$ (cf.~eq.~\req{vorticity_def})
  is color-coded and the flame position is indicated by solid
  white curves.
  \label{longterm_1}}
\end{figure*}

\begin{figure*}[p!]
\centerline{
\includegraphics[height=0.93 \textheight]{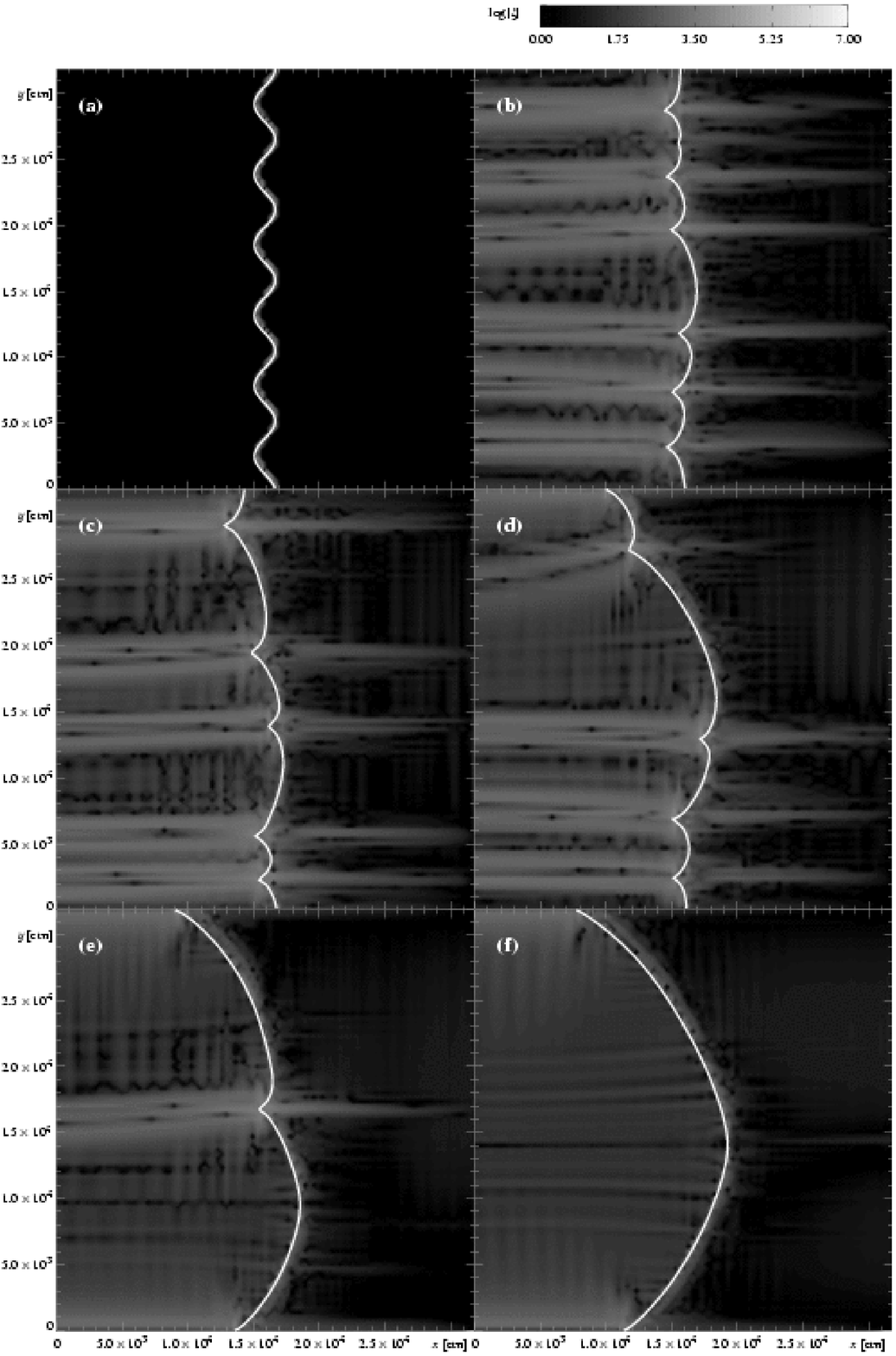}}
\caption{Flame propagation into quiescent fuel at $\rho_u = 5 \times
  10^7 \,\mathrm{g} \,\mathrm{cm}^{-3}$; resolution: $200 \times 200$ cells,
  \emph{reflecting} boundary conditions; Snapshots taken at
\textbf{(a)} 0, 
\textbf{(b)} 5,
\textbf{(c)} 7.5, 
\textbf{(d)} 10,
\textbf{(e)} 15, and 
\textbf{(f)} 30 growth times
  $\tau_\mathrm{LD} = \omega_\mathrm{LD}^{-1}$ of a
  perturbation with $\lambda=200 \Delta x$. The vorticity $\zeta$ (cf.~eq.~\req{vorticity_def})
  is color-coded and the flame position is indicated by solid
  white curves.
  \label{longterm_2}}
\end{figure*}

The numerical methods we apply are based on the work by
\citet{reinecke1999a} and have been described by
\citet{roepke2003a}. Therefore we will be brief in this and only
mention some keywords here refering to \cite {roepke2003a} for the
details. 

Since our simulations aim on the range around the Gibson scale, which is
well above the flame thickness for the fuel densities we are going to
apply, it is justified to model the flame as a \emph{discontinuity} between
unburnt and burnt states. This flame description ignores any internal
structure and therefore does not intrinsically provide the value of
laminar burning velocity. This also implies that, in a first approach,
hydrodynamics and flame propagation can be modeled in separate steps.

The hydrodynamics is modeled applying the
piecewise parabolic method (PPM)---a higher order Godunov scheme
developed by \citet{colella1984a}---in the \textsc{Prometheus} implementation
\citep{fryxell1989a}. The corresponding part of the code treats only the fluid
dynamics and does not account for the flame propagation. Since the
flame is modeled in the discontinuity approximation, its propagation
can be described applying the level-set method \citep{osher1988a}.
\citet{roepke2003a} could show that the key feature that enables us
to reproduce the theoretically anticipated flame evolution resulting
from the LD instability is a specific implementation of flame/flow
coupling, namely the in-cell reconstruction/flux-splitting
technique proposed by \citet{smiljanovski1997a}. It couples the flame
propagation accurately to the flow and therefore allows the
investigation of effects that originate from hydrodynamics, such as
the LD instability. Our implementation corresponds to what was termed
``complete implementation of the level-set method''
by \cite{reinecke1999a} with some minor changes
(cf.~\citealt{roepke2003a}). All simulations presented in the
following were performed in two spatial dimensions.

The basic approach is to study the flame evolution under the influence
of a small perturbation (which can be introduced by noise or---as in
our case---be initially imprinted to the flame structure). This
perturbation should grow due to the LD instability. In the nonlinear
regime, the formation of a cellular pattern should inhibit the further growth of the
perturbation and stabilize the flame. In a previous publication
\citep{roepke2003a} we demonstrated that this mechanism holds for
thermonuclear flames in degenerate matter.
From theoretical considerations and semi-analytical models (resulting
for instance from the
Sivashinsky equation, e.g.~\citealt{gutman1990a}), it can be expected that the general
features of the evolution of the flame front shape depend
significantly on the numerical setup used in the
simulations. Possible parameters that influence the flame evolution are
the overall geometry of flame propagation, 
the width of the computational domain compared to the length scale of
perturbations, the resolution, the boundary conditions, sources of
numerical noise (which is likely to be different when the simulation is
parallelized to a varying number of sub-processes), and the initial flame shape. Therefore care has
to be taken in choosing the specific setup depending on the questions
that are addressed by the simulations as well as in the conclusions drawn from
the simulation results. 

Regarding the overall flame geometry, two cases are commonly studied in
the literature: an on average planar flame geometry and a flame that is on
average circularly expanding. Although a naive approach would choose
the second scenario for the supernova explosion, it is probably not
the appropriate description of the flame propagation there. The flame
evolution on scales of
the star, where expansion effects are most pronounced, quickly becomes
dominated by the Rayleigh-Taylor instability. This leads to a flame
evolution completely diverging from a circular (or in three
dimensions spherical) geometry (cf.~\citealt{reinecke2002b}). It rather
proceeds in raising bubbles of burnt material. Nevertheless, it could be
argued that these structures again partly resemble a spherically expanding
geometry.
Although the case of a
circularly expanding flame reveals very interesting physical
effects (as repeated mode splitting of the cells resulting from
expansion effects and a possibly resulting fractalization of the flame
front, cf.~\citealt{blinnikov1996a}), we will not follow this approach
here for two reasons:
First, in the scope of our implementation it is
prohibitively expensive to follow a circular flame evolution for a
sufficiently long time and, second,
we aim on effects on scales
around the Gibson length, at which global expansion effects are
negligible. This has a technical advantage. It is much simpler to keep
an overall planar flame in the center of the domain by choosing a
comoving frame of reference. This enables us to study the long term
flame evolution without requiring large computational
domains. Thus it becomes possible simulate the
long-term flame evolution, as will be presented below.
Note that the choice of an overall planar flame geometry
determines a distinct flame evolution and possibly excludes the
mechanism of repeated mode splitting.

The influence of the width of the computational domain has already
been discussed by \citet{roepke2003a}. The simulations presented there
lead to the conclusion that in case of sufficient numerical
resolution
the flame stabilizes in a single domain-filling cusp-like structure
for periodic boundaries transverse to the direction of flame
propagation. However, with increasing domain width (corresponding to
a higher resolution in the simulations) this
fundamental structure became superimposed by a short-wavelength
cellular pattern. This is in accord with results from semi-analytical
studies \citep{gutman1990a}.

The discussion of the influence of the boundary conditions and the
initial shape of flame perturbations will be postponed to the next section.

The ``experimental setup'' applied our simulations is the following:
The spatial extent of the computational domain was set to
correspond to scales around the Gibson length and was fixed according
to the discussion above. 
The flame was initialized in the center of the domain in an on average planar
vertical shape with the unburnt material on the right hand side and
the burnt material on the left hand side, so that in laboratory frame
of reference the flame would propagate to the right.
The domain was set to be periodic in $y$-direction. On
the left boundary of the domain an outflow condition was enforced and
on the right boundary we imposed an inflow condition with the unburnt
material entering with the laminar burning velocity $s_l$. 
This would
yield a computational grid comoving with a planar flame. However, the LD
instability leads to a growth of the perturbation and therefore
increases the flame surface. According to eq.~\req{ueff_a_eq}, this causes an acceleration of the flame and therefore it is
necessary to
take additional measures in order to keep the mean position of the
flame centered in the domain. 
One possibility is to detect the mean
location of the flame and to simply shift the grid to keep the flame
in the center. This method is consistent with the  boundary conditions applied in
$x$-direction. The described method allows the study of the long term flame
evolution.
In order to induce the development of
perturbations, we usually imposed a sinusoidal perturbation on the
initial flame shape.

The state variables were set up with values for the burnt and
unburnt states obtained from (pseudo-)one-dimensional simulations
performed with the ``passive implementation'' of the level-set method
\citep{reinecke1999a}, imposing a value for the density of the
fuel. A compilation of the relevant setup values for
different fuel densities is given in Table \ref{values_tab}. 
Note that
frequently in this paper
fuel densities refer to a label of a specific set of
values rather than giving the accurate fuel density. The laminar
burning speed is calculated according to eq.~\req{TW_eq}.

\begin{table*}
\caption{Setup values for the simulations of the flame
  evolution.}
\label{values_tab}
$$
\begin{array}{p{0.11\linewidth}p{0.12\linewidth}p{0.12\linewidth}p{0.08\linewidth}
              p{0.08\linewidth}p{0.12\linewidth}p{0.12\linewidth}p{0.12\linewidth}}
\hline\hline
\noalign{\smallskip}
Label & \mbox{$\rho_u [\,\mathrm{g} \,\mathrm{cm}^{-3}]$} & \mbox{$\rho_b [\,\mathrm{g}
\,\mathrm{cm}^{-3}]$} & \mbox{$\mu$} & \mbox{$\mathit{At}$} & \mbox{$s_l [\,\mathrm{cm} \,\mathrm{s}^{-1}]$} (TW) &
\mbox{$e_{i,u} [\,\mathrm{erg} \,\mathrm{g}^{-1}]$}
& \mbox{$e_{i,b} [\,\mathrm{erg} \,\mathrm{g}^{-1}]$}\\
\noalign{\smallskip}
\hline
\noalign{\smallskip}
\mbox{$1 \times 10^7$} & \mbox{$9.969 \times 10^6$} & \mbox{$2.691 \times 10^6 $} & 3.70 &
0.575 & \mbox{$2.39 \times 10^5$} & \mbox{$1.619 \times 10^{17}$} & \mbox{$8.59 \times
10^{17}$}\\
\mbox{$1.25 \times 10^7$} & \mbox{$ 1.232 \times 10^7$} & \mbox{$3.830 \times 10^6 $} & 3.22 &
0.526 & \mbox{$2.86 \times 10^5$} & \mbox{$2.005 \times 10^{17}$} & \mbox{$8.99 \times
10^{17}$}\\
\mbox{$2.5 \times 10^7$} & \mbox{$ 2.484 \times 10^7$} & \mbox{$8.69 \times 10^6 $} & 2.86 &
0.482 & \mbox{$5.01 \times 10^5$} & \mbox{$2.56 \times 10^{17}$} & \mbox{$9.50 \times
10^{17}$}\\
\mbox{$5 \times 10^7$} & \mbox{$ 4.988 \times 10^7$} & \mbox{$2.071 \times 10^7 $} & 2.41 &
0.413 & \mbox{$8.74 \times 10^5$} & \mbox{$3.584 \times 10^{17}$} & \mbox{$1.051 \times
10^{18}$}\\
\mbox{$7.5 \times 10^7$} & \mbox{$ 7.50 \times 10^7$} & \mbox{$3.345 \times 10^7 $} & 2.24 &
0.383 & \mbox{$1.21 \times 10^6$} & \mbox{$4.29 \times 10^{17}$} & \mbox{$1.13 \times
10^{18}$}\\
\mbox{$1 \times 10^8$} & \mbox{$ 1.00 \times 10^8$} & \mbox{$4.789 \times 10^7 $} & 2.09 &
0.352 & \mbox{$1.53 \times 10^6$} & \mbox{$4.89 \times 10^{17}$} & \mbox{$1.182 \times
10^{18}$}\\
\mbox{$1 \times 10^9$} & \mbox{$ 1.00 \times 10^9$} & \mbox{$6.80 \times 10^8 $} & 1.47 &
0.190 & \mbox{$9.75 \times 10^6$} & \mbox{$1.27 \times 10^{18}$} & \mbox{$1.96 \times
10^{18}$}\\
\noalign{\smallskip}
\hline
\end{array}
$$
\end{table*}

\section{General features of flame evolution}
\label{quif_general_sec}

The flame evolution for the exemplary case of a fuel density of $5
\times 10 ^7 \,\mathrm{g} \,\mathrm{cm}^{-3}$ was described by
\citet{roepke2003a}. The results obtained in this reference 
were
\begin{itemize}
\item In the linear regime of flame evolution the initial perturbation
  grows due to the LD instability.
\item The formation of cusps at recesses of the flame front leads to a
  stabilization of the flame structure in a cellular shape in the
  nonlinear stage of evolution.
\item Different resolutions were tested. The flame stabilizes in a
  single domain-filling cusp-like structure for resolutions of $100
  \times 100$ cells and above.
\item In highly resolved simulations, the fundamental cusp-like
  structure became superimposed by a short-wavelength cellular
  pattern. This effect is in agreement with semianalytical results
  modeling the flame evolution with the Sivashinsky-equation
  \citep{gutman1990a}.
\item The growth of the perturbation amplitude in the linear regime is
  consistent with eq.~\ref{omega_ld} for sufficient resolution (above [200
  cells]$^2$ to [300 cells]$^2$).
\item The increase in flame surface area agrees reasonably well with
  the gain in the effective flame propagation speed.
\end{itemize}

In this section we extend this investigation by presenting
two simulations of the long-term flame evolution, focusing on the
nonlinear regime. 
The two simulations were carried out
on a regular Cartesian grid with $200\times200$ cells
with a cell width of $\Delta x = 160
\,\mathrm{cm}$. The fuel density was $5 \times 10^7 \,\mathrm{g} \,\mathrm{cm}^{-3}$. In
both cases we perturbed the flame front
initially in a sinusoidal way with a wavelength of 1/6 of the domain
width, but
we applied different boundaries transverse to the flame
propagation.

\begin{figure}[t]
\centerline{\resizebox{\hsize}{!}{
\includegraphics{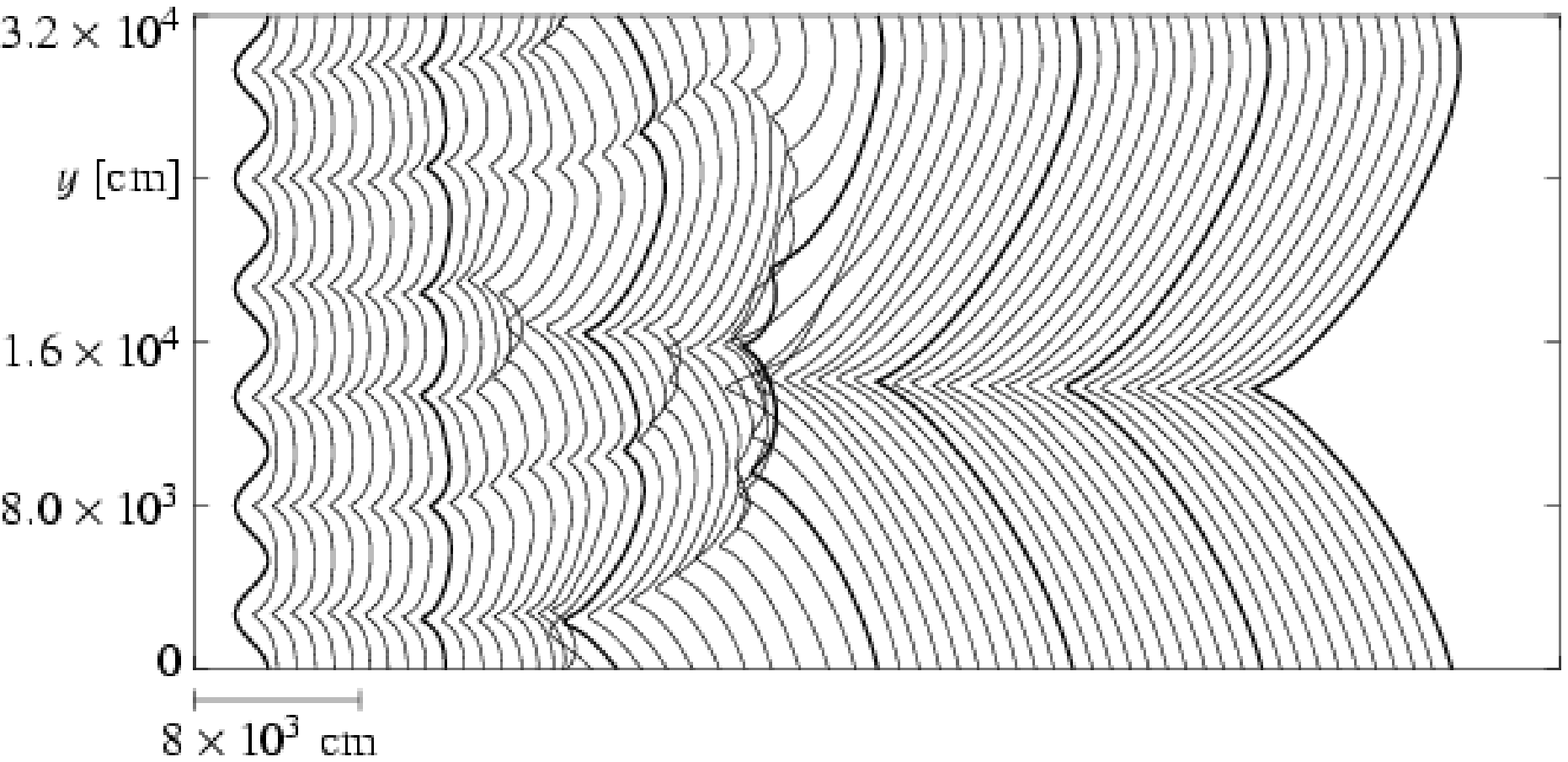}{\small(a)}}}
\centerline{\resizebox{\hsize}{!}{
\includegraphics{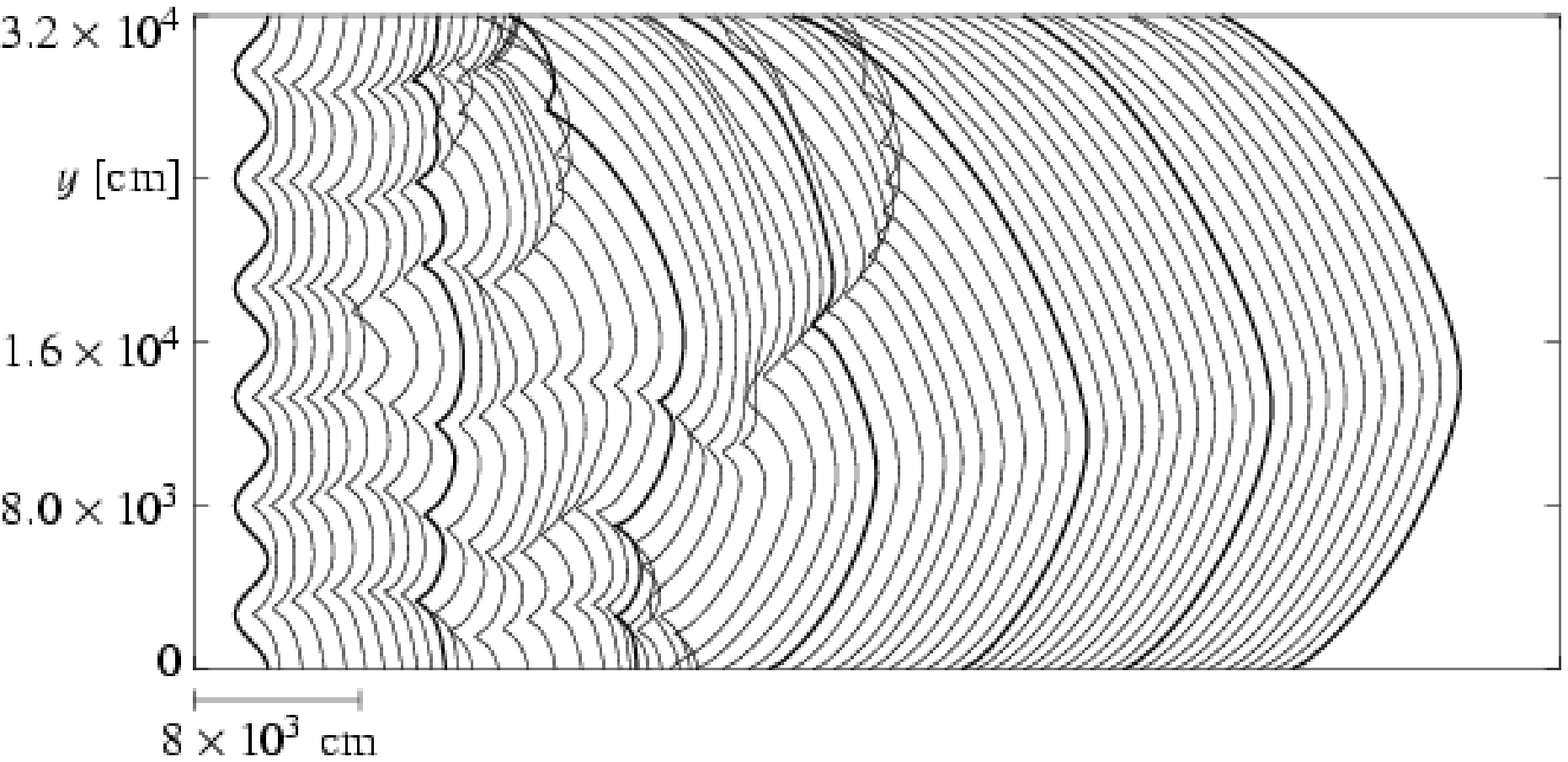}{\small(b)}}}
\caption{Evolution of the flame front for $\rho_u = 5 \times 10^7
  \,\mathrm{g} \,\mathrm{cm}^{-3}$ \textbf{(a)} with periodic boundaries in
  $y$-direction, \textbf{(b)} with reflecting boundaries in
  $y$-direction. Each contour marks a time step of 0.5
  $\tau_\mathrm{LD}(\lambda_\mathrm{pert}=3.2 \times 10^4\,\mathrm{cm})$.
  The contours are artificially shifted for better visibility and do
  not reflect the actual flame propagation velocity.
  \label{evo_longterm_fig}}
\end{figure}

Figure \ref{longterm_1} illustrates the flame evolution in case of periodic
boundary conditions in $y$-direction. The first snapshot shows the
initial setup with the sinusoidal perturbation imposed on the
flame. In the second snapshot the flame has already reached the
nonlinear regime and exhibits a cellular shape. Here, the onset of an
irregularity of the pattern is  apparent. This effect leads to a
merging of cells observable in the next three snapshots. 
Finally, the flame reaches a 
steady state which is a single domain-filling cusp. Note that the
snapshots in Fig.~\ref{longterm_1} were not taken at equal time
intervals. The final flame configuration was
followed for more than 10 growth times $\tau_{LD}=\omega_{LD}^{-1}$
corresponding to a perturbation of the width of the domain
in order to ensure its stability (cf.~Fig.~\ref{evo_longterm_fig}a).

In the snapshots of the flame evolution depicted in
Fig.~\ref{longterm_1} the vorticity of the flow field,
$\mbf{\zeta} = \mbf{\nabla} \times \mbf{v}$, is color-coded.
For our
two-dimensional simulations the (scalar) vorticity reads
\begin{equation}\label{vorticity_def}
  \zeta = \frac{\partial v_y}{\partial x} - \frac{\partial
    v_x}{\partial y}.
\end{equation}
Note that we color-code the logarithm of the
absolute value of the vorticity and therefore even slight
deviations from zero vorticity are visible. The flame produces vorticity in the
burnt material. This effect is especially strong behind
cusps. 
Analytical approximations of the flame evolution frequently make the assumption 
of a potential flow ahead of the flame. In contrast to that, we observe
the development of vorticity in the fuel region. This, however, is not
surprising, since the flame propagates subsonically and the regions
downstream and upstream of the flame are causally connected.

The merging
proceeds in a way that some of the cells grow and the smaller cells
disappear in the cusp between larger cells as can be clearly seen in
Fig.~\ref{evo_longterm_fig}a.
This effect forms the
basis of flame stability as explained by
\citet{zeldovich1980a}. There, the authors addressed the phenomenon
that flames in experiments are often observed to stabilize in large
cells. This, however, seems to be in contradiction with the result
from Landau's linear stability analysis (eq.~\req{omega_ld}), because
it predicts higher growth rates for perturbations with smaller
wavelengths, which thus should dominate the flame structure. Even taking
into account a modified dispersion relation
resulting from a finite flame thickness \citep{markstein1951a} cannot
explain this phenomenon. \citet{zeldovich1980a} argue that the stable
long-wavelength flame pattern is a result of the flow that establishes
upstream of the flame. A velocity component tangential to the flame is
directed toward the cusp and advects small perturbations into this
direction. A gradient in this tangential velocity stretches the
perturbation wavelength and thereby retards its growth. Finally,
the small perturbation disappears in the cusp (originally
\citet{zeldovich1980a} analyzed the flame propagation in tubes, but they
point out the similarity of this configuration with a cellular
flame). \citet{zeldovich1980a} give an analytical
description of the effect applying a WKB-like approximation. 

\begin{figure}[ht]
\resizebox{\hsize}{!}{
\includegraphics{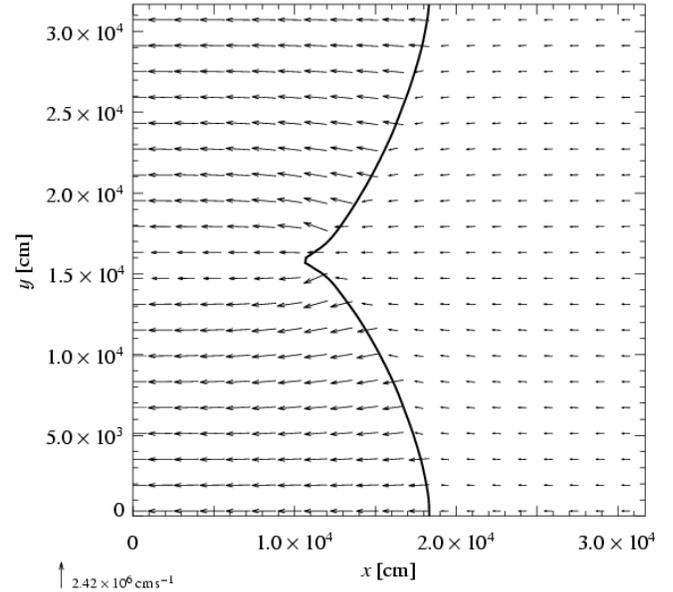}}
\caption{Velocity field in a simulation with a fuel density of $\rho_u=5 \times
  10^7\,\mathrm{g}\,\mathrm{cm}^{-3}$ and $100 \times 100$ cells resolution at
  $t= 0.09 \,\mathrm{s}$; solid curve indicates flame position.
  \label{velovect_fig}}
\end{figure}

The described
mechanism acts also in our numerical model, as was discussed by
\citet{roepke2003a} in
connection with the flame structure 
superposed by a short-wavelength cellular pattern for high-resolved
simulations. 
A close-up of the flow field in vicinity of a cusp taken
from one of our simulations (see Fig.~\ref{velovect_fig}) 
demonstrates
the ability of our
implementation to reproduce the effect.
The arrows represent the velocity field after the formation of a
cusp. Note that the streamlines converge toward
the cusp ahead of the front. This leads to the formation of a layer
upstream of the front in which the fluid velocity is directed toward
the cusp. Thus the required flow field providing stability of long-wavelength
patterns establishes in our simulations. In the downstream region, the
flow diverges from the cusp.

The flame evolution in case of \emph{reflecting boundary conditions}
in $y$-direction is shown in Fig.~\ref{longterm_2}. The mechanism of
cellular stabilization acts similar to the case of periodic
boundaries. However, the alignment of the steady-state of the flame
shape is different. Whereas the periodic case developed a cusp in the
center of the domain, in the reflecting case the crest of the pattern
centers in the domain and the recesses of the front form a structure
similar to a ``half-cusp'' at the boundaries. This case is
analogous to the configuration studied by \citet{zeldovich1980a}.

Another aspect of these two simulations should be
noted. \citet{roepke2003a} studied the evolution of a flame that was
initially perturbed by only one single domain-filling mode. This was
done in order to be more efficient in the investigation of the linear
growth due to the LD instability and it was argued that a single
domain-filling structure establishes independently of the exact shape
of the initial perturbation. This is reconfirmed by the presented
simulations in accord with expectations from numerical simulations of
the Sivashinsky-equation (e.g.~\citealt{gutman1990a}) and from the
pole-decomposition solutions of that equation \citep{thual1985a}.
In both the simulation with periodic boundaries
(cf.~Fig.~\ref{longterm_1})  and the simulation 
with reflecting boundary conditions (see Fig.~\ref{longterm_2}) the
cells merge. The steady state of the flame resulting from this process
is a single domain-filling cell.

\section{Flame stability at different fuel densities}
\label{quif_fuel_sec}

Further numerical experiments addressed the flame stability at
different fuel densities. With lower fuel density, the Atwood
number---defined as $\mathit{At} = |\rho_u - \rho_b|/(\rho_u+\rho_b)$---
and, equivalently, the density contrast $\mu$ over the flame front
increase (cf.~Table \ref{values_tab}). At the same time the laminar burning velocity of the flame
decreases. In the following, tests of the flame
stability at a variety of fuel densities will be presented. These use the setup
introduced in Sect.~\ref{setup_sec} with periodic boundaries in the
direction transverse to the flame propagation. We again refer to Table
\ref{values_tab} for a compilation of the setup values. The resolution
chosen for this study was $200\times 200$ cells. From the simulations
presented in the previous section, it follows that an initial flame
perturbation with a wavelength corresponding to the width of the
computational domain is sufficient in order to see the flame stabilize
in a universal steady state. In the case under consideration one
expects a single centered domain-filling cusp-like
structure.  Although the restriction to an initial domain-filling
perturbation wavelength may appear rather artificial, it avoids the
time-consuming cell merging process before the final steady-state flame
structure is reached.

\begin{figure}[t]
  \centerline{\resizebox{\hsize}{!}{
  \includegraphics{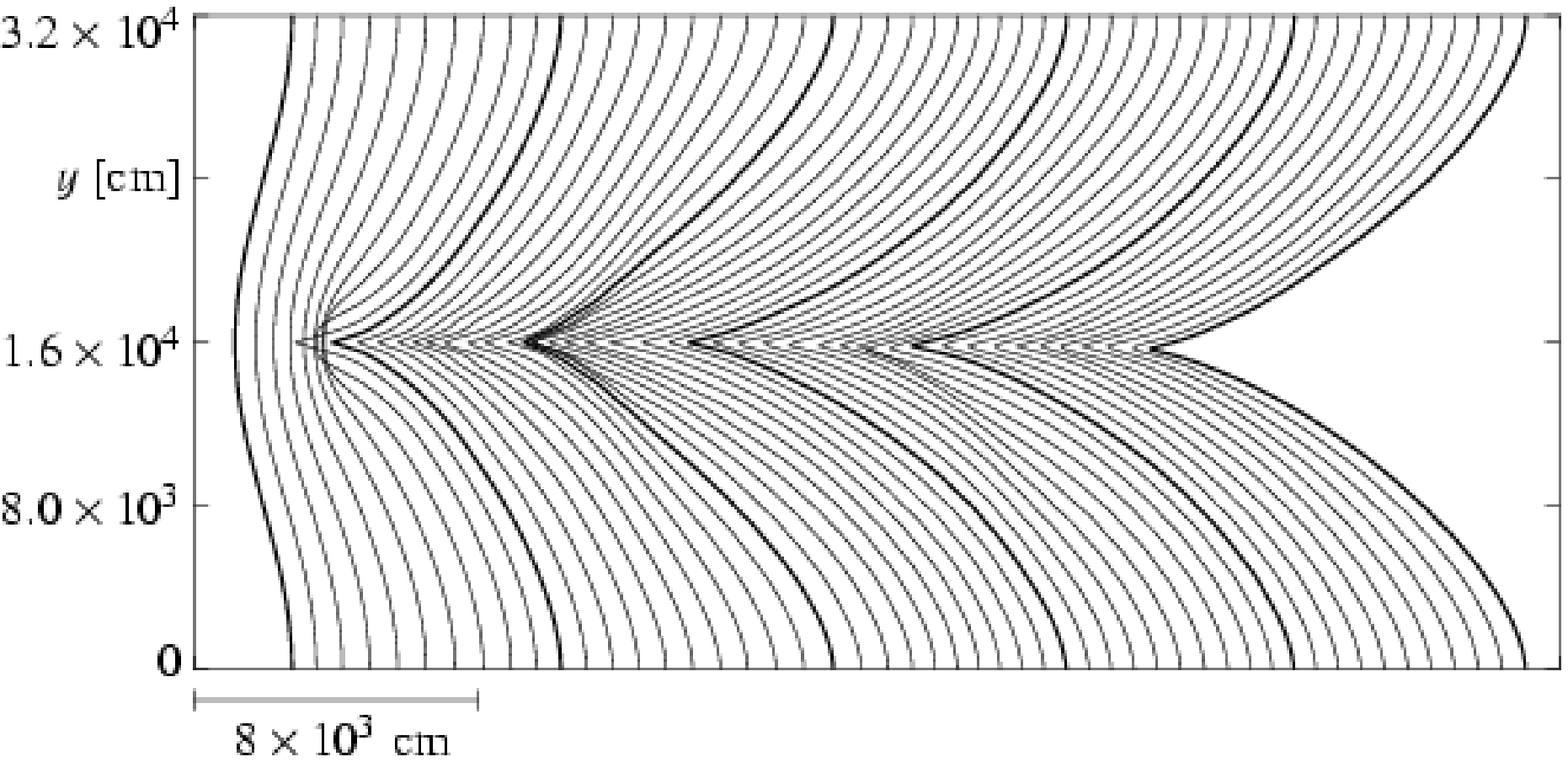}{\small(a)}}}
  \centerline{\resizebox{\hsize}{!}{
  \includegraphics{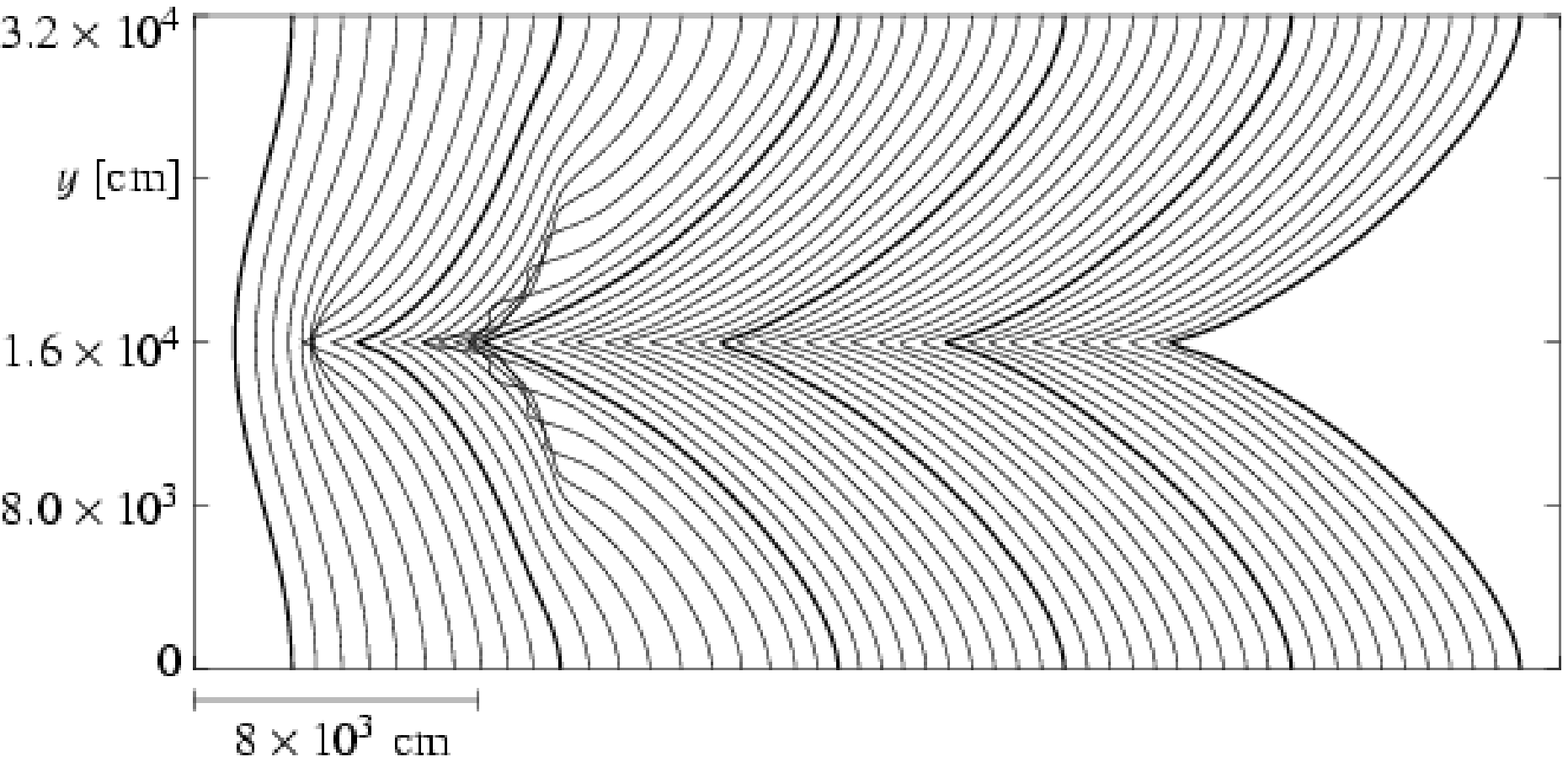}{\small(b)}}}
  \caption{Evolution of the flame front at \textbf{(a)} $\rho_u=2.5\times 10^7
    \,\mathrm{g}\,\mathrm{cm}^{-3}$ and \textbf{(b)} $\rho_u=5\times 10^7
    \,\mathrm{g}\,\mathrm{cm}^{-3}$. The
    contours mark evolution steps of 0.2 $\tau_\mathrm{LD}$. Every
    10th conture is highlighted bold for better visibility.
  \label{evo_dens_1_fig}}
\end{figure}

Figure~\ref{evo_dens_1_fig}
shows the temporal evolution of the flame front for fuel densities of
$2.5\times 10^7 \,\mathrm{g}\,\mathrm{cm}^{-3}$ and $5\times 10^7
\,\mathrm{g}\,\mathrm{cm}^{-3}$. In both cases the flame stabilizes in a
single domain-filling cusp-like structure. Note, that the evolution
time has been normalized to the growth time of the LD instability $\tau_\mathrm{LD}
= \omega_\mathrm{LD}^{-1}$
corresponding to the initial perturbation wavenumber. This ensures comparability between the
different simulations. As in Fig.~\ref{evo_longterm_fig}, the spacing between individual contours is, however,
artificial since the original simulations were performed in the frame
of reference comoving with the flame. It has been chosen in a way that
the contours fill the plot window and does not reflect flame
propagation. 

\begin{figure}[t]
  \centerline{\resizebox{\hsize}{!}{
  \includegraphics{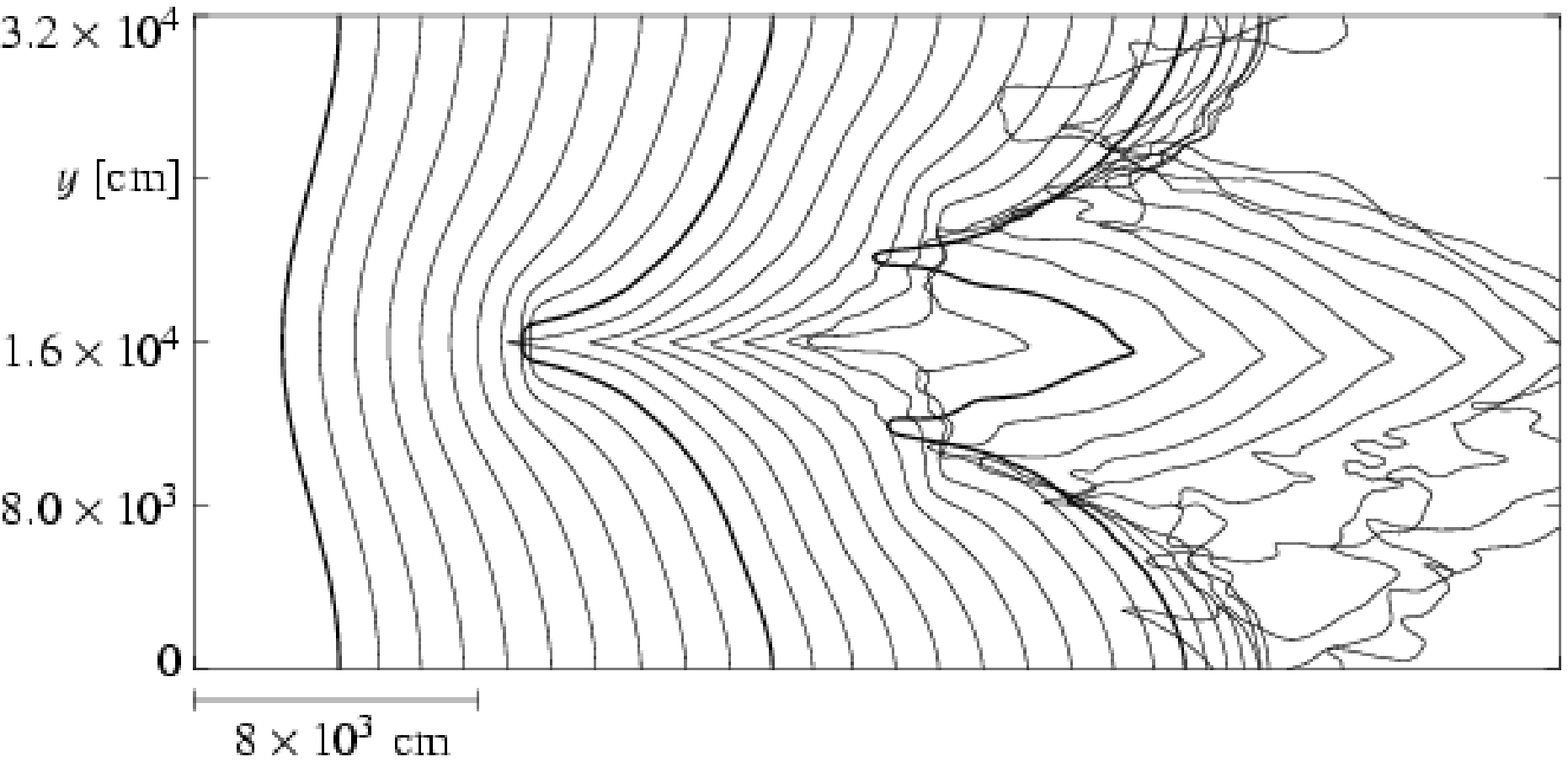}{\small(a)}}}
  \centerline{\resizebox{\hsize}{!}{
  \includegraphics{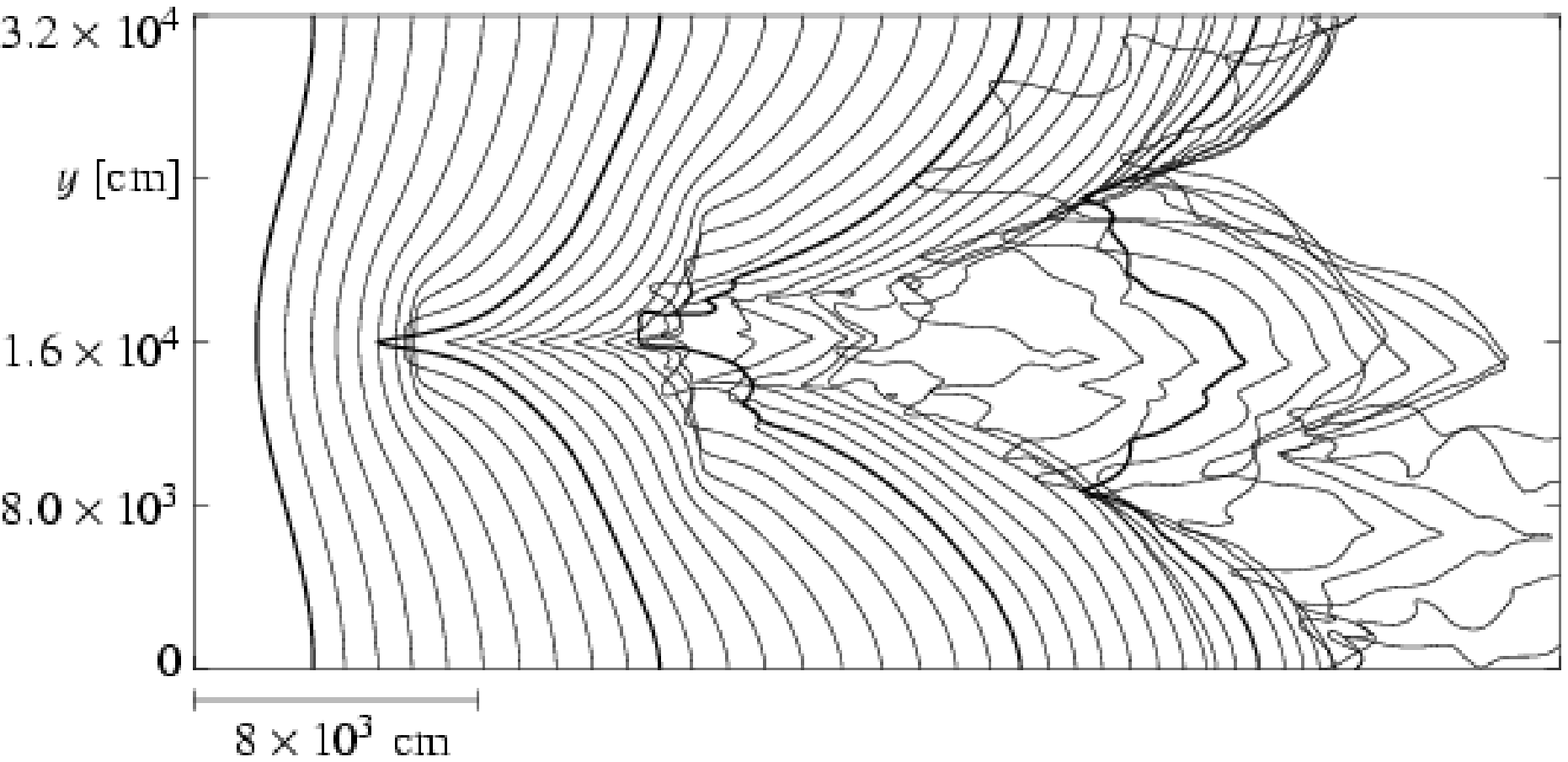}{\small(b)}}}
  \caption{Evolution of the flame front at \textbf{(a)} $\rho_u=1\times 10^7
    \,\mathrm{g}\,\mathrm{cm}^{-3}$ and \textbf{(b)} $\rho_u=1.25 \times 10^7
    \,\mathrm{g}\,\mathrm{cm}^{-3}$. The
    contours mark evolution steps of 0.2 $\tau_\mathrm{LD}$. Every
    10th conture is highlighted bold for better visibility.
  \label{evo_dens_2_fig}}
\end{figure}

The outcome from simulations with even lower fuel densities reveals a
completely different flame evolution. Figure~\ref{evo_dens_2_fig} shows the
flame for fuel densities of  $1\times 10^7 \,\mathrm{g}\,\mathrm{cm}^{-3}$ and $1.25\times 10^7
\,\mathrm{g}\,\mathrm{cm}^{-3}$. Although the initial flame evolution
resembles that of $\rho_u \approx 5 \times 10^7
\,\mathrm{g}\,\mathrm{cm}^{-3}$, the forming cusp lacks long-term
stability. As can be seen from the plots, the stable structure breaks
up from the cusps outward and the flame subsequently evolves in an irregular pattern.
The origin of these effects is most likely numerical.
For a fuel density of $1\times 10^7 \,\mathrm{g}\,\mathrm{cm}^{-3}$ a
simulation run with $600 \times 600$ cells resolution did not result
in a disruption of the cusp. Thus,
a high grid
resolution is required in order to describe a stable cusp properly at low fuel densities
($\lesssim\,$$1.25 \times 10^7 \,\mathrm{g}\,\mathrm{cm}^{-3}$). 

\begin{figure}[ht]
  \centerline{\resizebox{\hsize}{!}{
  \includegraphics{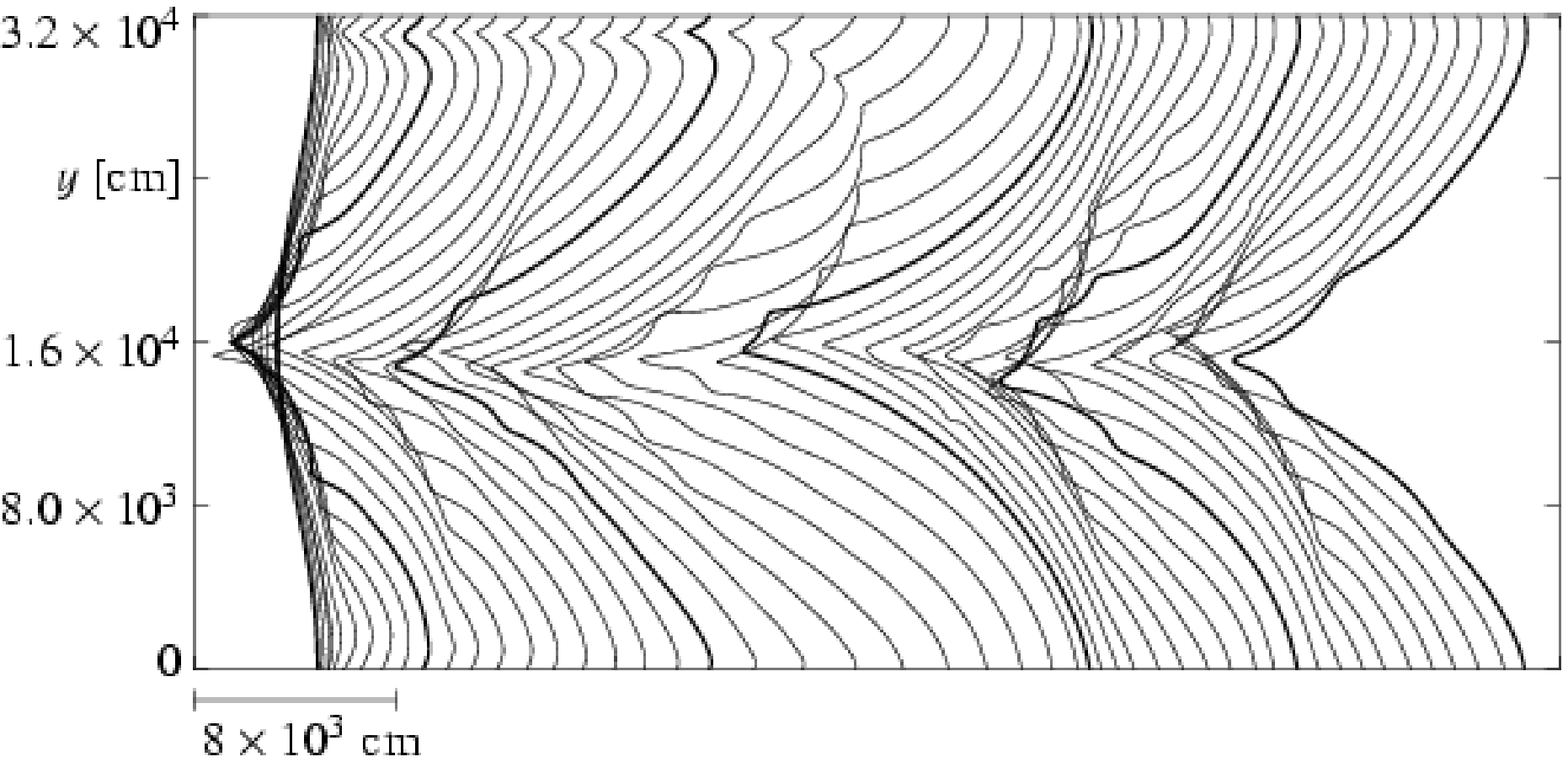}{\small(a)}}}
  \centerline{\resizebox{\hsize}{!}{
  \includegraphics{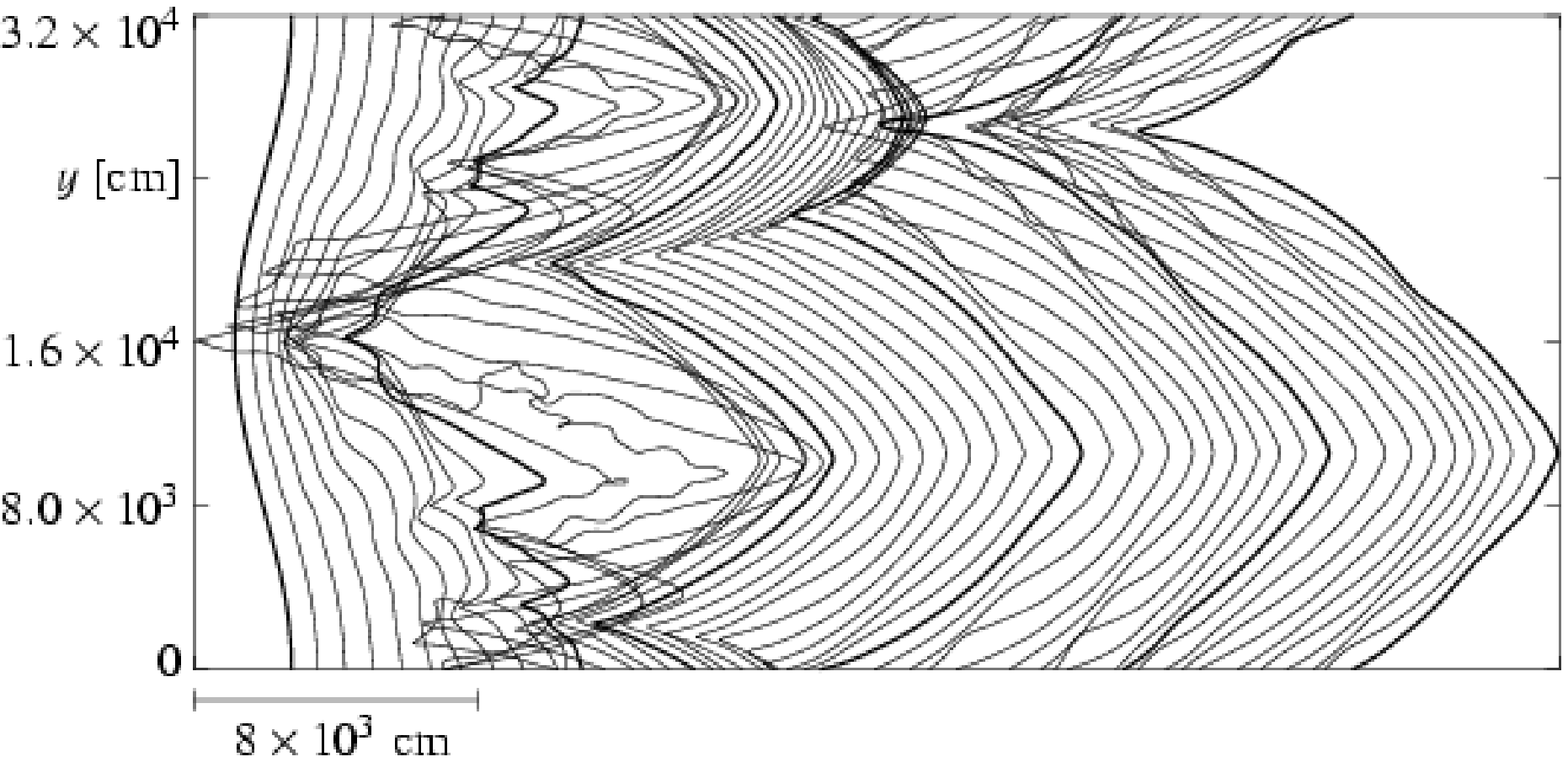}{\small(b)}}}
  \caption{Evolution of the flame front at \textbf{(a)} $\rho_u=7.5\times 10^7
    \,\mathrm{g}\,\mathrm{cm}^{-3}$ and \textbf{(b)} $\rho_u=1\times 10^8
    \,\mathrm{g}\,\mathrm{cm}^{-3}$. The
    contours mark evolution steps of 0.2 $\tau_\mathrm{LD}$. Every
    10th conture is highlighted bold for better visibility.
  \label{evo_dens_3_fig}}
\end{figure}

A rather unexpected result is obtained from simulations for fuel densities around
 $1\times 10^8 \,\mathrm{g}\,\mathrm{cm}^{-3}$. 
Here, the flame does not propagate in a
stable manner in the first stages of the flame evolution. 
This is illustrated by
Fig.~\ref{evo_dens_3_fig}. Small-wavelength perturbations superpose
the initial flame shape shortly after the beginning of the
simulation. These appear to dominate the flame evolution for a while
(see in particular
Fig.~\ref{evo_dens_3_fig}b) but then the flame stabilizes
in the single-cell structure. This implies that (i) the flame
structure is less stable against small-wavelength
perturbations at these fuel densities and (ii) the mechanism for a
stabilization in a preferred long-wavelength pattern as analytically
predicted (see the discussion in Sect.~\ref{quif_general_sec}) does
finally stabilize the flame evolution.

\begin{figure}[ht]
  \centerline{\resizebox{\hsize}{!}{
  \includegraphics{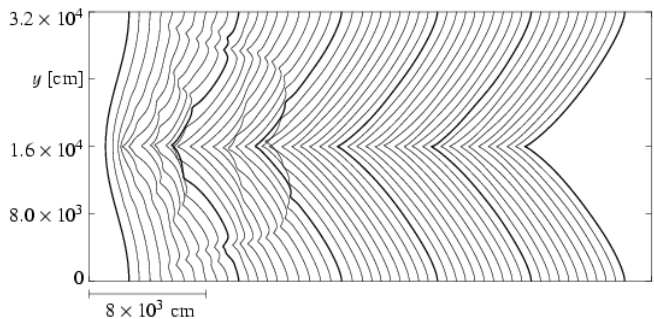}}}
  \caption{Evolution of the flame front at $\rho_u=1\times 10^9
    \,\mathrm{g}\,\mathrm{cm}^{-3}$. The
    contours mark evolution steps of 0.2 $\tau_\mathrm{LD}$. Every
    10th conture is highlighted bold for better visibility.
  \label{evo_dens_4_fig}}
\end{figure}

The initial flame destabilization becomes less pronounced in flame
evolution at even higher densities. Figure~\ref{evo_dens_4_fig} shows an
example with $\rho_u = 1\times 10^9 \,\mathrm{g}\,\mathrm{cm}^{-3}$.

\begin{figure*}[t]
  \centerline{
  \includegraphics[width = 0.8\linewidth]{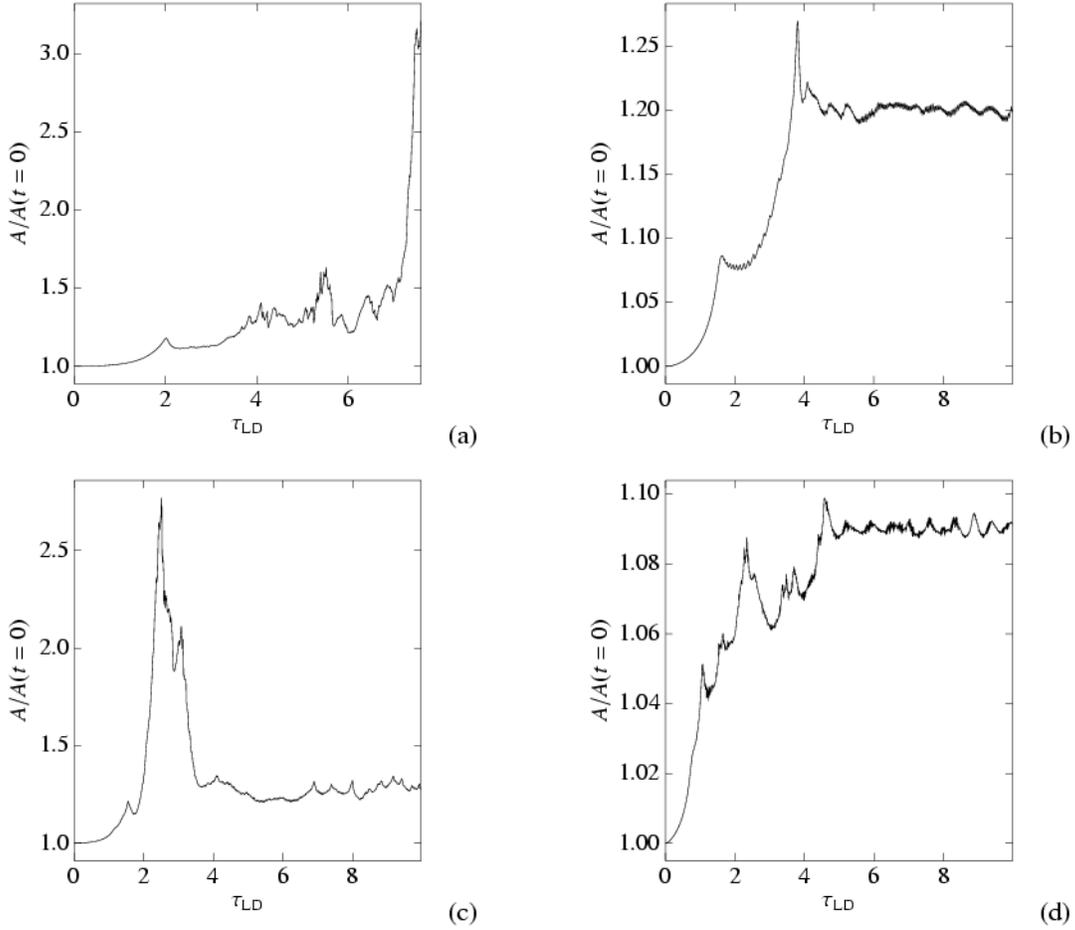}}
  \caption{Increase in flame surface area at fuel densities of
  \textbf{(a)} $\rho_u=1.25 \times 10^7 \,\mathrm{g}\,\mathrm{cm}^{-3}$,
  \textbf{(b)} $\rho_u=5 \times 10^7 \,\mathrm{g}\,\mathrm{cm}^{-3}$,
  \textbf{(c)} $\rho_u=1 \times 10^8 \,\mathrm{g}\,\mathrm{cm}^{-3}$,
  \textbf{(d)} $\rho_u=1 \times 10^9 \,\mathrm{g}\,\mathrm{cm}^{-3}$.
  \label{area_fig}}
\end{figure*}

We measured the flame acceleration in these simulations via the
increase in flame surface area (cf.~eq.~\req{ueff_a_eq}). 
This method
has been discussed by \citet{roepke2003a}. Although there are some
minor differences compared to measuring the effective flame propagation
velocity directly (most important, eq.~\req{ueff_a_eq} strictly holds
only in case of a flame idealized as a discontinuity, which is fulfilled
to a reasonable degree in our simulations, but probably not quite
exactly, cf.~\citealt{roepke2003a}), this approach can be
expected to yield rather reliable results.
Some
examples of the temporal evolution of the flame surface area are
plotted in Fig.~\ref{area_fig}. This figure illustrates the different
cases of flame evolution. While the simulations of low fuel densities
(Fig.~\ref{area_fig}a) fail to converge, the flame reaches a
steady-state propagation velocity for higher fuel densities. The
initial destabilization of the flame pattern and final stabilization
for fuel densities around $10^8 \,\mathrm{g} \,\mathrm{cm}^{-3}$ is apparent in
the evolution of the corresponding flame surface area
(Fig.~\ref{area_fig}c).

\begin{table}
  \caption{Flame propagation velocities in the cellular regime.}
  \label{vel_tab}
  $$
    \begin{array}{p{0.3\linewidth}p{0.3\linewidth}p{0.35\linewidth}}
      \hline\hline
      \noalign{\smallskip}
      \mbox{$\rho_u$ [$\,\mathrm{g} \,\mathrm{cm}^{-3}$]} & \mbox{$v_\mathrm{cell} / s_l$}\linebreak (measured) &
      \mbox{$v_\mathrm{cell} / s_l$}\linebreak (according to eq.~(\ref{vel_inc_eq}))\\
      \noalign{\smallskip}
      \hline
      \noalign{\smallskip}
      \mbox{2.5 $\hspace{0.5em}\times 10^7$} & 1.23 & 1.17 \\
      \mbox{5 $\hspace{1.2em}\times 10^7$} & 1.20 & 1.14 \\
      \mbox{7.5 $\hspace{0.5em}\times 10^7$} & 1.25 & 1.13 \\
      \mbox{1 $\hspace{1.28em}\times 10^8$} & 1.30 & 1.11 \\
      \mbox{1.25 $\hspace{0.01em} \times 10^8$} & 1.35 & 1.10 \\
      \mbox{1 $\hspace{1.28em}\times 10^9$} & 1.09 & 1.04 \\
      \noalign{\smallskip}
      \hline
    \end{array}
  $$
\end{table}

Table \ref{vel_tab} compares the final steady-state velocities of the
cellular flames
to the theoretical predictions from
eqs.~(\ref{vel_inc_eq}). These equations, of
course, ignore any
structure superimposed to the front the flame front and additionally
assumes a parabolic
shape of the smooth cells.
Both effects obscure the expected trent of decreasing
$v_\mathrm{cell}/s_l$ with increasing fuel density. This can be
clearly seen from Figs.~\ref{evo_dens_1_fig}--\ref{evo_dens_4_fig},
where the deviations are most pronounced for fuel densities around
$10^8 \,\mathrm{g} \,\mathrm{cm}^{-3}$.
Therefore the measurements agree only in the
order of magnitude with the theoretical values. This indicates that
ignoring the superimposed
small-wavelength pattern is a too simple approach and underestimates
the actual flame surface.

\section{Conclusions}

The simulations presented in Sects.~\ref{quif_general_sec} and
\ref{quif_fuel_sec} provide a deeper insight into the dynamics of
thermonuclear flame fronts under conditions of SN Ia explosions. One
fundamental assumption was that the flame propagates into \emph{quiescent}
fuel.

In Sect.~\ref{quif_general_sec} we addressed the question of the influence of the
boundary conditions and the shape of the initial flame perturbation on
the long-term flame evolution.
The
conclusions that can be drawn from our simulations are:
\begin{enumerate}
\item
In the given
setup, the initial perturbation shape is not retained. Although at
first it develops into a cellular structure of the same
wavelength in the nonlinear regime, the flame later shows
the tendency to align in the shape of a single domain-filling
cell. The transition to this steady-state pattern happens by growth of
preferred cells from the initial perturbation and the disappearance of
smaller cells in newly formed cusps. This effect gives rise to a
``merging process'' of cells, 
consistent
with numerical solutions \citep{gutman1990a} and with pole
decomposition analysis
\citep{thual1985a} of the Sivashinsky-equation \citep{Sivashinsky1977a}.
\item
The merging of the small cells proceeds smoothly without drastic
effects on the flame shape. In particular it does not lead to a loss
of flame stability.
\item
The cases of reflecting and periodic boundary conditions differ in the
alignment of the ultimately developed cell. In the case of
periodic boundary conditions the cusp develops at the center of the
domain and the crest tends toward the boundary. In principle no
specific alignment should be preferred for periodic boundaries. This is
not true for reflecting boundary
conditions. Here, the crest centers in the
domain, similar to the simulation with adiabatic boundaries presented
by \citet{gutman1990a}.
\end{enumerate}

These results support the generality of the simulations presented
by \citet{roepke2003a} and were used to set-up the study of the influence
of the fuel density on the flame stability that was presented in
Sect.~\ref{quif_fuel_sec}. Here, the general features of the flame evolution
resemble those of \citet{roepke2003a} and Sect.~\ref{quif_general_sec}
of the present paper. However, some peculiarities were observed. At
very low fuel densities ($\lesssim 1.25 \times 10^7\,\mathrm{g} \,\mathrm{cm}^{-3}$), the
cusp-like flame shape breaks up. It was argued that this is a
numerical effect. Nevertheless, if the reason for the disruption
of the stable flame pattern is numerical noise, then one could conclude
that flames at lower fuel densities are likely to become more
sensitive to velocity fluctuations in the fuel region. These, however,
can be expected to be present in realistic scenarios of SN Ia
explosions. Therefore we will address that question in a subsequent study.

A destabilization of the flame structure shortly
after initialization was observed at densities around $10^8 \,\mathrm{g}
\,\mathrm{cm}^{-3}$. However, after $\sim$$2 \tau_\mathrm{LD}$ the flame stabilized again
forming the cusp-like steady-state structure. Thus, we conclude that
the flame tends to stabilize in a cellular pattern in the parameter
space explored in our simulations.

The implications of the presented investigation for large-scale SN Ia
models can be summarized as follows: 

(i) The flame is expected to
propagate in a stable way at least down to fuel densities of around $10^7
\,\mathrm{g} \,\mathrm{cm}^{-3}$. This corroborates fundamental assumptions of
large-scale SN Ia models (e.g.~\citealt{reinecke2002d}) which rely on
a smooth flame evolution on unresolved scales. We disagree with previous simulations
by \citet{niemeyer1995a}, who reported on a possible loss of flame
stability at a fuel density of $5 \times 10^7 \,\mathrm{g} \,\mathrm{cm}^{-3}$.
Therefore, a self-turbulization and subsequent drastic acceleration of
the flame seems unlikely---at least in the case of propagation into
quiescent fuel. No convincing indications for \emph{active turbulent combustion} 
\citep{niemeyer1997b,kerstein1996a}) were observed in our simulations. However, flame
propagation into a turbulent flow field may reveal completely
different flame structures, since one expects strong enough turbulent
eddies to disrupt the cellular flame pattern. The arguments above
provide strong hints into this direction.

(ii) Large-scale SN Ia simulations as performed by \citet{reinecke2002d}
apply a subgrid-scale model to determine the contribution from unresolved
scales to the effective flame propagation velocity. However, as a lower
cut-off they use the laminar burning velocity $s_l$. On the
basis of the performed simulations we suggest a slightly higher value
in order to take into account the cellular structure of the flame
below the Gibson scale. This may have impact on the nucleosynthesis
yields from SN Ia models and will be addressed in a forthcoming study.

\begin{acknowledgements}
This work was supported in part by the European Research Training
Network ``The Physics of Type Ia Supernova Explosions'' under contract
HPRN-CT-2002-00303 and by the DFG Priority Research Program ``Analysis
and Numerics for Conservation Laws'' under contract HI 534/3.
A pleasant atmosphere to prepare this publication was provided at the
workshop ``Thermonuclear Supernovae and Cosmology'' at the ECT*,
Trento, Italy. 
We would like to thank M.~Reinecke,
S.~Blinnikov, and W.~Schmidt for stimulating discussions.
The numerical simulations were performed on an IBM Regatta
system at the computer center of the Max Planck Society in Garching. 
\end{acknowledgements}

\end{document}